\documentclass[aps,prb,preprint]{revtex4}

\usepackage{graphics}

\begin{document}

\title{Thermodynamic properties of the SO(5)\ model of high-$T_{c}$
superconductivity}
\author{T. A. Zaleski}
\author{T. K. Kope\'{c}}
\affiliation{Institute of Low Temperature and Structure Research, Polish Academy of\\
Sciences, P.O. Box 1410, 50-950 Wroc\l aw, Poland}

\begin{abstract}
In this paper we present calculations of thermodynamic functions within
Zhang's SO(5) quantum rotor theory of high-$T_{c}$ superconductivity. Using
the spherical approach for the three-dimensional quantum rotors we derieved
explicit analytical formulas for entropy and specific heat related to the
lattice version of the SO(5)\ nonlinear quantum-sigma model. We present the
temperature dependence of these quantities for various settings of relevant
control parameters (quantum fluctuations, chemical potential). We find our
results in overal qualitative agreement with basic thermodynamics of high-$%
T_{c}$ cuprates.
\end{abstract}

\maketitle

\section{Introduction}

The theory unifying antiferromagnetism (AF) and superconductivity (SC) to
describe global phase diagram of high-$T_{c}$ superconductors was recently
proposed by Zhang. \cite{ZhangScience} In this approach, based on symmetry
principles, a three-dimensional order parameter (the staggered
magnetization) describing the AF phase and a complex order parameter (with
two real components), describing a spin singlet d-wave SC phase are grouped
in five-component vector called a \textquotedblleft
superspin\textquotedblright . The SO(3) symmetry of spin rotations (which is
spontaneously broken in the AF phase) and the electro-magnetic SO(2)
invariance (whose breaking defines SC phase) along with well-defined AF to
SC and vice versa rotation operators form SO(5) symmetry. In the Zhang's
theory both ordered phase arise once SO(5)\ is spontaneously broken and the
competition between antiferromagnetism and superconductivity is related to
direction of the \textquotedblleft superspin\textquotedblright in the
five-dimensional space. The low energy dynamics of the system is determined
in terms of the Goldstone bosons and their interactions specified by the
SO(5) symmetry. The kinetic energy of the system is that of a SO(5)\ rigid
rotor and the system is described by a SO(5) non-linear quantum $\sigma $
model (NLQ$\sigma $M). The SO(5) quantum rotor model offers a
Landau-Ginzburg-like (LG)\ approach for the high-$T_{c}$ problem.\ However, it goes
much beyond the traditional LG\ theory, since it captures dynamics. While
the SO(5)\ symmetry was originally proposed in the context of an effective
field-theory description of the high-$T_{c}$ superconductors, its prediction
can also be tested within microscopic models. \cite%
{MicroscopicModels1,MicroscopicModels2,MicroscopicModels3,MicroscopicModels4,MicroscopicModels5,Arrigoni}
For example, numerical evidence for approximate SO(5)\ symmetry of the
Hubbard model came out from exact diagonalization of small sized clusters. 
\cite{Meixner} The global features of the phase diagram deduced from SO(5)
theory based on a spherical quantum rotors \cite{PRB1} agree qualitatively
with the general topology of the observed phase diagram of high-$T_{c}$
superconductors. The quantitative investigation of the quantum critical
point scenario within the concept of the SO(5) group, \textit{e.g.} the
scaling of the contribution to the electrical resistivity due spin
fluctuations showed linear resistivity dependence on temperature for
increasing quantum fluctuation -- being a hallmark example of anomalous
properties in cuprate materials. \cite{PRL} The systematic studies of
magnetic properties of the SO(5) theory showed that the theory yields a
qualitative scenario for the evolution of magnetic behavior, which is
consistent with experiments. \cite{PRB2} It qualitatively explains the
results of experimental measurements (notably the NMR relaxation rates) with
correct predictions of behavior of uniform spin susceptibility in high
temperatures. Also the energy dependence of the momentum-integrated dynamic
spin susceptibility show features, which are in qualitative agreement with
experimental findings.

Thermal fluctuations are pronounced in the high-$T_{c}$ superconductors due
to number of reasons. The carrier density is rather small, the anisotropy is
large and the critical temperature is high. It turns out that deviations
from the mean-field behavior are present in the specific heat $C$ at the
superconducting transition temperature $T_{c}$. In the mean-field BCS
theory, a second order transition with a jump in a specific heat at $T_{c}$
takes place. In contrary, in most high-$T_{c}$ superconductors thermal
fluctuations seem to restore common behavior.

The aim of the present paper is to study quantitatively basic thermodynamic
functions resulting from the SO(5) theory, thereby substantiating this
theoretical framework. Our study may also provide a useful diagnostic tool
for testing the basic principles of SO(5) theory by comparing the
quantitative predictions (\textit{e.g.}, specific heat) with the outcome of
the relevant experiments.

The outline of the reminder of the paper is as follows. In Section II we
begin by setting up the quantum SO(5) Hamiltonian and the corresponding
Lagrangian. In Section III we find closed forms of various thermodynamic
functions. We calculate free energy, entropy and specific heat. Finally, in
Section IV we summarize the conclusions to be drawn from out work.

\section{Hamiltonian and the effective Lagrangian}

We consider the low-energy Hamiltonian of superspins placed in the nodes of
a discrete three dimensional simple cubic (3DSC) lattice, 
\begin{eqnarray}
H &=&\frac{1}{2u}\sum_{i}\sum_{\mu <\nu }L_{i}^{\mu \nu }L_{i}^{\mu \nu
}-\sum_{i<j}J_{ij}\mathbf{n}_{i}\cdot \mathbf{n}_{j}+  \nonumber \\
&&-V\left( \mathbf{n}_{i}\right) -2\mu \sum_{i}L_{i}^{15}\text{.}
\label{Eq_Hamiltonian}
\end{eqnarray}

Indices $i$ and $j$ number lattice sites running from $1$ to $N$ - the total
number of sites, while $\mu $,$\nu =1,...,5$ denote superspin $\mathbf{n}%
_{i}=\left( n_{1},n_{2},n_{3},n_{4},n_{5}\right) _{i}$ components ($\mathbf{n%
}_{AF,i}=\left( n_{2},n_{3},n_{4}\right) _{i}$ refers to antiferromagnetic
and $\mathbf{n}_{SC,i}=\left( n_{1},n_{5}\right) _{i}$ superconducting
order, respectively). The superspin components are mutually commuting
(according to Zhang's formulation) and their values are restricted by the
rigidity constraint $\mathbf{n}_{i}^{2}=1$.

The first part of the equation (\ref{Eq_Hamiltonian}) is the kinetic energy
of the system (being simply that of a SO(5) rigid rotor), where 
\begin{equation}
L_{i}^{\mu \nu }=n_{\mu i}p_{\nu i}-n_{\nu i}p_{\mu i}
\end{equation}%
are generators of Lie SO(5) algebra (expressed by total charge $L_{i}^{15}$,
spin and so-called \textquotedblleft $\pi $\textquotedblright operators), 
$p_{\mu i}$ are momenta conjugated to respective superspin components: 
\begin{eqnarray}
&&p_{\mu i}=i\frac{\partial }{\partial n_{\mu i}},  \nonumber \\
&&\left[ n_{\mu },p_{\nu }\right] =i\delta _{\mu \nu }
\end{eqnarray}%
and parameter $u$ measures the kinetic energy of the rotors (an analog of moment of inertia).

The second part of the Hamiltonian is the inter-superspin interaction energy
with $J$ being the stiffness in the charge and spin channel. In the 3DSC
lattice, $J$ is nonvanishing for the nearest neighbors and its Fourier
transform
\begin{equation}
J_{\mathbf{q}}=\frac{1}{N}\sum_{\mathbf{R}_{i}}J\left( \mathbf{R}_{i}\right)
e^{-i\mathbf{R}_{i}\cdot \mathbf{q}}  \label{FourierTransform}
\end{equation}%
is simply 
\begin{equation}
J_{\mathbf{q}}/J=\cos q_{x}+\cos q_{y}+\cos q_{z}.  \label{Eq_Jq}
\end{equation}%
For convenience, we will further introduce the density of states 
\begin{equation}
\rho \left( \xi \right) =\frac{1}{N}\sum_{\mathbf{q}}\delta \left( \xi -J_{%
\mathbf{q}}/J\right) \text{,}  \label{Eq_DOS}
\end{equation}%
which for 3DSC\cite{3d} lattice reads: 
\begin{eqnarray}
\rho \left( \xi \right)  &=&\frac{1}{\pi ^{3}}\int_{\max (-1,-2-\xi )}^{\min
(1,2-\xi )}\frac{dy}{\sqrt{1-y^{2}}}\times   \nonumber \\
&&\times \mathbf{K}\left[ \sqrt{1-\left( \frac{\xi +y}{2}\right) ^{2}}\right]
\Theta \left( 3-\left\vert \xi \right\vert \right) \text{,}
\end{eqnarray}%
where $\mathbf{K}\left( x\right) $ is the elliptic integral of the first
kind and $\Theta \left( x\right) $ is the step function \cite%
{EllipticFunction}.

The last two parts of the equation (\ref{Eq_Hamiltonian}) provide symmetry
SO(5) breaking terms. In the result of their interplay, the system favours
either the \textquotedblleft easy plane\textquotedblright in the
superconducting (SC) space $\left( n_{1},n_{5}\right) $, or an
\textquotedblleft easy sphere\textquotedblright in the antiferromagnetic
(AF) space $\left( n_{2},n_{3},n_{4}\right) $. The first of two terms is
defined as: 
\begin{equation}
V\left( \mathbf{n}_{i}\right) =\frac{w}{2}\sum_{i}\left(
n_{2i}^{2}+n_{3i}^{2}+n_{4i}^{2}\right) ,
\end{equation}%
with the anisotropy constant $w$, which positive value favours AF state. The
second term contains a charge operator $L_{i}^{15}$, whose expectation value
yields the doping concentration and the chemical potential $\mu $ (measured
from the half-filling), whose positive values favour SC state.

The partition function $Z=Tre^{-H/k_{B}T}$ is expressed using the functional
integral in the Matsubara \textquotedblleft imaginary time\textquotedblright 
$\tau $ formulation \cite{PRB1} ($0\leq \tau \leq 1/k_{B}T\equiv \beta $,
with $T$ being the temperature). We obtain:

\begin{eqnarray}
Z &=&\int \prod_{i}\left[ D\mathbf{n}_{i}\right] \int \prod_{i}\left[ \frac{D%
\mathbf{p}_{i}}{2\pi }\right] \delta \left( 1-\mathbf{n}_{i}^{2}\right)
\delta \left( \mathbf{n}_{i}\cdot \mathbf{p}_{i}\right) \times  \nonumber \\
&&\times e^{-\int_{0}^{\beta }d\tau \left[ i\mathbf{p}\left( \tau \right)
\cdot \frac{d}{d\tau }\mathbf{n}\left( \tau \right) +H\left( \mathbf{n},%
\mathbf{p}\right) \right] }=  \nonumber \\
&=&\int \prod_{i}\left[ D\mathbf{n}_{i}\right] \delta \left( 1-\mathbf{n}%
_{i}^{2}\right) e^{-\int_{0}^{\beta }d\tau \mathcal{L}\left( \mathbf{n}%
\right) },  \label{Eq_PartitionFunction}
\end{eqnarray}
with $\mathcal{L}$ being the Lagrangian:

\begin{eqnarray}
\mathcal{L}\left( \mathbf{n}\right) &=&\frac{1}{2}\sum_{i}\left[ u\left( 
\frac{\partial \mathbf{n}_{SC}}{\partial \tau }\right) ^{2}+u\left( \frac{%
\partial \mathbf{n}_{AF}}{\partial \tau }\right) ^{2}+\right.  \nonumber \\
&&\left. -4u\mu ^{2}\mathbf{n}_{SC}^{2}+4iu\mu \left( \frac{\partial n_{1}}{%
\partial \tau }n_{5}-\frac{\partial n_{5}}{\partial \tau }n_{1}\right) %
\right] +  \nonumber \\
&&-\sum_{i<j}J_{ij}\mathbf{n}_{i}\cdot \mathbf{n}_{j}-\frac{w}{2}%
\sum_{i}\left( n_{2i}^{2}+n_{3i}^{2}+n_{4i}^{2}\right) \text{.}
\label{Eq_Lagrangian}
\end{eqnarray}

The problem can be solved \textit{exactly} in terms of the spherical model%
\cite{Sphere}. To accommodate this we notice that the superspin rigidity
constraint ($\mathbf{n}_{i}^{2}=1$ ) implies that a weaker condition also
holds, namely: 
\begin{equation}
\sum_{i=1}^{N}\mathbf{n}_{i}^{2}=N\text{.}  \label{SphericalConstraint}
\end{equation}

Therefore, the superspin components $\mathbf{n}_{i}\left( \tau \right) $
must be treated as $c$-number fields, which satisfy the quantum periodic
boundary condition $\mathbf{n}_{i}\left( \beta \right) =\mathbf{n}_{i}\left(
0\right) $ and are taken as \textit{continuous} variables, i.e., $-\infty <%
\mathbf{n}_{i}\left( \tau \right) <\infty $, but constrained (on average,
due to Eq. (\ref{SphericalConstraint})) to have unit length.

This introduces the Lagrange multiplier $\lambda \left( \tau \right) $
adding an additional quadratic term (in $\mathbf{n}_{i}$ fields) to the
Lagrangian (\ref{Eq_Lagrangian}). Fourier transform $\mathbf{n}(\mathbf{k}%
,\omega _{\ell })$ of the superspin components 
\begin{equation}
\mathbf{n}_{i}\left( \tau \right) =\frac{1}{\beta N}\sum_{\mathbf{k}%
}\sum_{\ell =-\infty }^{\infty }\mathbf{n}\left( \mathbf{k},\omega
_{\ell}\right) e^{-i\left( \omega _{\ell}\tau -\mathbf{k\cdot r}_{i}\right) }
\end{equation}
introduces the Matsubara (Bose) frequencies $\omega _{\ell}=2\pi \ell/\beta $
($\ell =0,\pm 1,\pm 2,...$).

Using the Eq. (\ref{Eq_PartitionFunction}), the partition function can be
written in the form: 
\begin{equation}
Z=\int \frac{d\lambda }{2\pi i}e^{-N\phi \left( \lambda \right) }\text{,}
\end{equation}
where the function $\phi \left( \lambda \right) $ is defined as: 
\begin{eqnarray}
\phi \left( \lambda \right) &=&-\int_{0}^{\beta }d\tau \lambda \left( \tau
\right) -\frac{1}{N}\ln \int \prod_{i}\left[ D\mathbf{n}_{i}\right] \times 
\nonumber \\
&&\times \exp \left[ -\sum_{i}\int_{0}^{\beta }d\tau \left( \mathbf{n}%
_{i}^{2}\lambda \left( \tau \right) -\mathcal{L}\left[ \mathbf{n}\right]
\right) \right] \text{.}  \label{PhiFunction}
\end{eqnarray}

The exact value of the partition function can be found in the thermodynamic
limit ($N\rightarrow \infty $), when the method of steepest descents is
exact and the saddle point $\lambda \left( \tau \right) =\lambda _{0}$
satisfies the condition: 
\begin{equation}
\left. \frac{\delta \phi \left( \lambda \right) }{\delta \lambda \left( \tau
\right) }\right| _{\lambda =\lambda _{0}}=0\text{.}  \label{SaddlePoint}
\end{equation}
At criticality, corresponding order parameters susceptibilities becomes
infinite and corresponding Lagrange multipliers are: 
\begin{eqnarray}
\lambda _{0}^{AF} &=&\frac{1}{2}J_{\mathbf{k}=0}+\frac{w}{2},  \nonumber \\
\lambda _{0}^{SC} &=&\frac{1}{2}J_{\mathbf{k}=0}+2\chi \mu ^{2},
\label{LambdaZero}
\end{eqnarray}
for AF\ and SC critical lines, respectively. Furthermore, using the
spherical condition (\ref{SphericalConstraint}) and the values (\ref%
{LambdaZero}), we finally arrive at the expression for the critical lines
separating AF, SC and QD (quantum disordered) states (for more specific
description these calculations see, Ref. \onlinecite{PRB1,PRB2}).

\section{Thermodynamic functions}

\subsection{Free energy}

The free energy is defined as $f=-\left( \beta N\right) ^{-1}\ln Z=\left(
\beta \right) ^{-1}\phi \left( \lambda _{0}\right) $. Using the formula (\ref%
{PhiFunction}), we obtain:

\begin{widetext}%
\begin{eqnarray}
f &=&-\lambda +\frac{3%
}{2\beta N}\sum_{\mathbf{k},\ell}\ln \left[ 2\lambda -J_{\mathbf{k}}+u\omega
_{\ell}^{2}-w\right] + \\
&&+\frac{1}{2\beta N}\sum_{\mathbf{k},\ell}\ln \left[ 2\lambda -J_{\mathbf{k}%
}+u\left( \omega _{\ell}+2i\mu \right) ^{2}\right] +\frac{1}{2\beta N}\sum_{%
\mathbf{k},\ell}\ln \left[ 2\lambda -J_{\mathbf{k}}+u\left( \omega _{\ell}-2i\mu
\right) ^{2}\right]\text{.}  \nonumber
\end{eqnarray}
\end{widetext}

After performing the summation over Matsubara's frequencies, we obtain the
free energy: 
\begin{eqnarray}
f &=&-\lambda +\frac{1}{\beta }\int_{-\infty }^{\infty }\rho \left( \xi
\right) d\xi \left\{ 3\ln 2\sinh \left( \frac{\beta }{2}\sqrt{\frac{2\lambda
-J\xi -w}{u}}\right) \right. +  \nonumber \\
&&+\ln 2\sinh \left[ \frac{\beta }{2}\left( \sqrt{\frac{2\lambda -J\xi }{u}}%
-2\mu \right) \right]  \label{FreeEnergy_Final} \\
&&+\left. \ln 2\sinh \left[ \frac{\beta }{2}\left( \sqrt{\frac{2\lambda
-J\xi }{u}}+2\mu \right) \right] \right\} .  \nonumber
\end{eqnarray}

\subsection{Entropy}

The entropy is defined as %
$S=k_{B}\beta ^{2}\frac{\partial f}{\partial \beta }$. %
Using the formula (\ref{FreeEnergy_Final}) we obtain:

\begin{widetext}%
\begin{eqnarray}
S\left( \beta \right) &=&\frac{k_{B}}{2}\int_{-\infty }^{\infty }\rho \left(
\xi \right) d\xi \left\{ 3\left[ \beta A\left( \xi \right) \coth \frac{\beta 
}{2}A\left( \xi \right) -2\ln 2\sinh \frac{\beta }{2}A\left( \xi \right) %
\right] +\right. \\
&&+ \left[ \beta B_{-}\left( \xi \right) \coth \frac{\beta }{2}%
B_{-}\left( \xi \right) -2\ln 2\sinh \frac{\beta }{2}B_{-}\left( \xi \right) %
\right] \nonumber \\&&+\left. 
\left[ \beta B_{+}\left( \xi \right) \coth \frac{\beta }{2}%
B_{+}\left( \xi \right) -2\ln 2\sinh \frac{\beta }{2}B_{+}\left( \xi \right) %
\right] \right\} ,  \nonumber
\end{eqnarray}
\end{widetext}where:

\begin{eqnarray}
A\left( \xi \right) &=&\sqrt{\frac{2\lambda -J\xi -w}{u}},  \nonumber \\
B_{-}\left( \xi \right) &=&\sqrt{\frac{2\lambda -J\xi }{u}}-2\mu,  \nonumber
\\
B_{+}\left( \xi \right) &=&\sqrt{\frac{2\lambda -J\xi }{u}}+2\mu.
\label{ABBDefs}
\end{eqnarray}

\begin{figure}[htb]
\scalebox{0.6}{\includegraphics{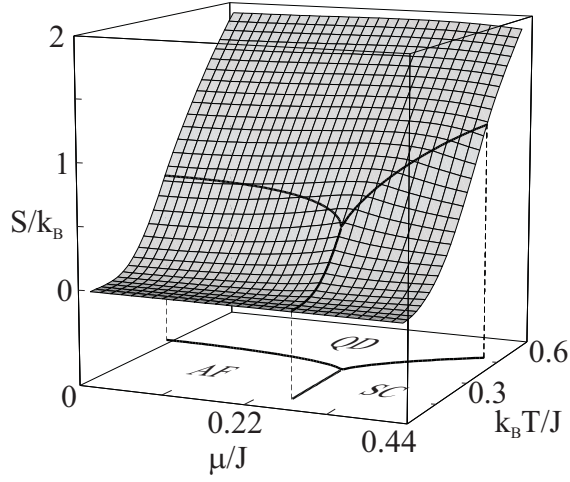}}
\caption{Plot of the entropy $S$ vs. chemical potential $\protect\mu /J$ and
temperature $k_{B}T/J$ for fixed $uJ=3$ and $w/J=1$. Solid lines indicate
the projection of the $\protect\mu -T$ phase diagram.}
\label{Fig_Entropy}
\end{figure}

\begin{figure}[htb]
\scalebox{0.5}{\includegraphics{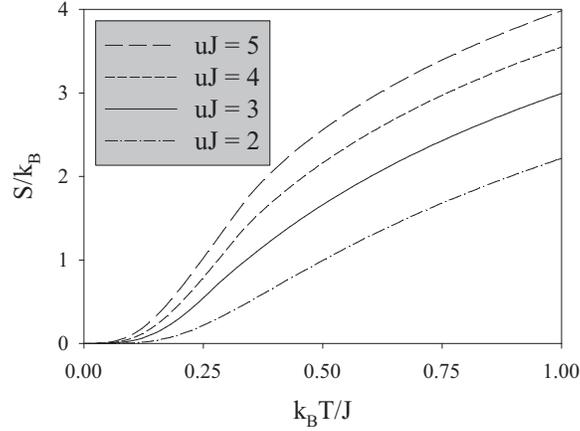}}
\caption{Plot of the entropy $S$ vs. temperature $k_{B}T/J$ for $w/J=1$, $%
\protect\mu /J=0.2$ and different values of $uJ$, as indicated in the inset.}
\label{Fig_Entropy2}
\end{figure}

The dependence of the entropy on temperature and chemical-potential is shown
on Fig. \ref{Fig_Entropy}. Starting from $T=0$, the entropy increases in any
ordered phase (AF or SC) until reaching $T_{c}$ (or $T_{N}$). The further
increase is slower, but saturation in higher temperatures is not observed.
The absolute value of the entropy is lower for higher quantum fluctuation
(see, Fig. \ref{Fig_Entropy2}). We find obtained results in qualitative
agreement with experimentally measured properties of high-$T_{c}$
superconductors (\textit{e.g.} for Bi2212 compound, see Ref. %
\onlinecite{Loram_Entropy}).

\subsection{Specific heat}

The specific heat at constant volume is defined: 
\begin{eqnarray}
C &=&-k_{B}\beta ^{2}\frac{\partial ^{2}}{\partial \beta ^{2}}\left( \beta
f\right) =-k_{B}\beta ^{2}\left\{ 2\frac{\partial f}{\partial \beta }+\beta 
\frac{\partial ^{2}f}{\partial \beta ^{2}}\right.  \nonumber \\
&&\left. +\beta \frac{d\lambda }{d\beta }\left[ \frac{\partial ^{2}f}{%
\partial \lambda ^{2}}\frac{d\lambda }{d\beta }+2\frac{\partial ^{2}f}{%
\partial \lambda \partial \beta }\right] \right\} .
\end{eqnarray}
The derivative $d\lambda /d\beta $ can be found from the saddle-point
condition (\ref{SaddlePoint}): 
\begin{equation}
\left. \frac{\partial f\left[ \lambda \left( \beta \right) ,\beta \right] }{%
\partial \lambda }\right| _{\lambda =\lambda _{0}}=0 .
\end{equation}
Explicitly, we obtain: 
\begin{equation}
\frac{d\lambda }{d\beta }=-\frac{\frac{\partial ^{2}f}{\partial \lambda
\partial \beta }}{\frac{\partial ^{2}f}{\partial \lambda ^{2}}} .
\end{equation}
The specific heat: 
\begin{equation}
C=-k_{B}\beta ^{2}\left[ 2\frac{\partial f}{\partial \beta }+\beta \frac{%
\partial ^{2}f}{\partial \beta ^{2}}+\beta \frac{d\lambda }{d\beta }\frac{%
\partial ^{2}f}{\partial \lambda \partial \beta }\right] .
\end{equation}
Using the formula (\ref{FreeEnergy_Final}) we obtain: 
\begin{widetext}%

\begin{eqnarray}
C &=&\frac{k_{B}\beta ^{2}}{4}\int_{-\infty }^{\infty }\rho \left( \xi
\right) d\xi \times  \nonumber \\
&&\times \left\{ 3A^{2}\left( \xi \right) \sinh ^{-2}\frac{\beta }{2}A\left(
\xi \right) +B_{-}^{2}\left( \xi \right) \sinh ^{-2}\frac{\beta }{2}%
B_{-}\left( \xi \right) +B_{+}^{2}\left( \xi \right) \sinh ^{-2}\frac{\beta 
}{2}B_{+}\left( \xi \right) \right\} +  \nonumber \\
&&+\frac{k_{B}\beta ^{3}}{4u}\frac{d\lambda }{d\beta }\int_{-\infty
}^{\infty }\rho \left( \xi \right) d\xi \times  \nonumber \\
&&\times \left\{ 3\sinh ^{-2}\frac{\beta }{2}A\left( \xi \right) +\frac{%
B_{-}\left( \xi \right) }{C\left( \xi \right) }\sinh ^{-2}\frac{\beta }{2}%
B_{-}\left( \xi \right) +\frac{B_{+}\left( \xi \right) }{C\left( \xi \right) 
}\sinh ^{-2}\frac{\beta }{2}B_{+}\left( \xi \right) \right\} ,
\end{eqnarray}
where $A\left( \xi \right) $, $B_{-}\left( \xi \right) $ and $B_{+}\left(
\xi \right) $ are defined by formulas (\ref{ABBDefs}), 
\begin{equation}
C\left( \xi \right) =\sqrt{\frac{2\lambda -J_{\mathbf{k}}}{u}} ,
\end{equation}
and 
\begin{eqnarray}
\frac{{d\lambda }}{{d\beta }} &=& - \frac{u}{2}\int_{ - \infty }^{ + \infty
} {\rho \left( \xi \right)d\xi \left\{ {3\sinh ^{ - 2} \frac{\beta }{2}%
A\left( \xi \right) + \frac{{B_ - \left( \xi \right)}}{{C\left( \xi \right)}}%
\sinh ^{ - 2} \frac{\beta } {2}B_ - \left( \xi \right) + \frac{{B_ + \left(
\xi \right)}}{{C\left( \xi \right)}} \sinh ^{ - 2} \frac{\beta }{2}B_ +
\left( \xi \right)} \right\}} /  \nonumber \\
&&\int_{ - \infty }^{ + \infty } {\rho \left( \xi \right)d\xi \left\{ {\frac{%
\beta } {{2A^{2}\left( \xi \right)}}\sinh ^{ - 2} \frac{\beta }{2}A\left( \xi
\right) + \frac{{\coth \frac{\beta }{2}A\left( \xi \right)}}{{A^3 \left(
\xi \right)}} + \frac{\beta }{{2C^{2}\left( \xi \right)}}\sinh ^{ - 2} \frac{%
\beta }{2}B_ - \left( \xi \right)} \right.} + \nonumber \\
&&\left. {\ + \frac{{\coth \frac{\beta }{2}B_ - \left( \xi \right)}}{{C^3
\left( \xi \right)}} + \frac{\beta }{{2C^{2}\left( \xi \right)}}\sinh ^{ - 2} 
\frac{\beta }{2}B_ + \left( \xi \right) + \frac{{\coth \frac{\beta }{2}B_ +
\left( \xi \right)}}{{C^3 \left( \xi \right)}}} \right\}.
\end{eqnarray}
\end{widetext}

\begin{figure}[htb]
\scalebox{0.5}{\includegraphics{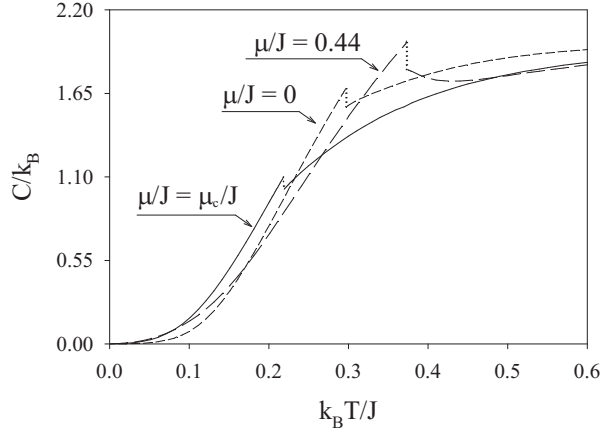}}
\caption{Specific heat vs. temperature $k_{B}T/J$ for $w/J=1$, $uJ=3$ and
various values of chemical-potential, as indicated on the figure.}
\label{Fig_SpecHeat}
\end{figure}

The temperature dependence of the specific is presented in Fig. \ref%
{Fig_SpecHeat}. The low temperature behavior of $C(T)$ may be approximated
by $C(T)\sim T^{3}$ for $\mu /J=0$ and $C(T)\sim T^{2.5}$ for $\mu /J=0.44$.
For higher temperatures (but still below transition temperature) the linear
behavior of the specific heat is observed. Reaching the critical temperature
($T_{c}$ or $T_{N}$), the specific heat experiences a finite jump (implying
the value $\alpha = 0$ for the critical exponent of the specific heat). For
higher temperatures, saturation is observed.

\section{Summary and final remarks}

In conclusion, we have calculated entropy and specific heat dependence on
temperature and various other parameters using the unified theory of
antiferromagnetism and superconductivity proposed for the high-$T_{c}$
cuprates by Zhang and based on the SO(5)\ symmetry between antiferromagnetic
and superconducting states. The theory of yields a qualitative scenario for
the evolution of thermodynamic functions behavior, which is consistent with
experiments. Most experimental work on the specific heat in the high-$T_{c}$
superconductors have concentrated on yttrium compound (Y-123). \cite{Junod1,
Junod2, Junod3} Optimally doped Y-123 does not show a jump in the specific
heat, but $\lambda $-peak at $T_{c}$. However, the shape for overdoped Y-123
is intermediate between a BCS step and a $\lambda $-type transition.
Furthermore, optimally doped Bi-2212 shows a symmetric anomaly (intermediate
between $\lambda $-peak and finite jump). Experimentally, the specific heat
is not very sensitive to the critical exponent $\alpha $ and one can
ascertain that $|\alpha |\ll 1$ for Y-123 compounds. However, the result of
the present work ($\alpha =0$) agrees with the critical behavior of the
3D-XY model. Finally, checking the validity of basic principles of the SO(5)
theory, by comparing parameters discussed here with relevant, obtained from
calculations on microscopic models of high-$T_{c}$ superconductors is still
called for.


\begin{thebibliography}{99}
\bibitem{3d} Our approach is not restricted to the three dimensional cubic
lattice and can be easily accommodated to virtually any other lattice (by
using the proper density of states function).
\end{thebibliography}
\end{document}